\documentclass[prd,aps,nofootinbib]{revtex4}
\usepackage{amsmath}
\usepackage{graphicx}
\usepackage{amsmath}
\usepackage{amssymb}
\usepackage{latexsym}
\usepackage{color}

\newcommand{\GB}{\mathcal{G}}
\newcommand{\de}{\mathrm{d}}
\newcommand{\rew}{\frac{16\pi}{M_P^2} w}

\DeclareMathOperator*{\Tr}{tr}

\begin{document}

\title{Scalar mode propagation in modified gravity with a scalar field}

\author{Antonio De Felice} \email{antonio.defelice@uclouvain.be}
\affiliation{Theoretical and Mathematical Physics Group, Centre for
  Particle Physics and Phenomenology, Louvain University, 2 Chemin du
  Cyclotron, 1348 Louvain-la-Neuve (Belgium)}

\author{Teruaki Suyama}
\email{teruaki.suyama@uclouvain.be}
\affiliation{Theoretical and Mathematical Physics Group, Centre for
  Particle Physics and Phenomenology, Louvain University, 2 Chemin du
  Cyclotron, 1348 Louvain-la-Neuve (Belgium)}

\date{\today}

\begin{abstract}
 We study the propagation of the scalar modes around a Friedmann-Lema\^\i tre-Robertson-Walker universe for general modifications of gravity in the presence of a real scalar field. In general, there will be two propagating scalar perturbation fields, which will have in total four degrees of freedom. Two of these degrees will have a superluminal propagation---with $k$-dependent speed of propagation---whereas the other two will have the speed of light. Therefore, the scalar degrees of freedom do not modify the general feature of modified gravity models: the appearance of modes whose frequency depends on the second power of the modulus of the wave vector. Constraints are given and special cases are discussed.
\end{abstract}

\maketitle

\section{Introduction}

In \cite{DeFelice:2009ak}, the scalar cosmological perturbation theory
around a general Friedmann-Lema\^\i tre-Robertson-Walker (FLRW)
universe was studied for those general modified gravity models (MGM)
\cite{2005PhRvD..71f3513C,Nunez:2004ts,Nojiri:2007te,Navarro:2005gh,Navarro:2005da,Mena:2005ta,DeFelice:2006pg}
whose action is given by
\begin{equation}
S=\frac{M_P^2}{16\pi} \int \de^4x\sqrt{-g}~f(R,\GB), \label{actionvacuum}
\end{equation}
where $f(R,\GB)$ is a general function, $R$ is Ricci scalar, and $\GB$
is the so-called Gauss-Bonnet term defined by
\begin{equation}
\GB\equiv R^2-4R_{\mu \nu} R^{\mu \nu}+R_{\mu \nu \alpha \beta} R^{\mu \nu \alpha \beta}.
\end{equation}
It was found that, in general, the propagation equation for the metric
perturbation was acquiring a term proportional to $k^4$, where $k$ is
the length of the comoving wavenumber vector $k \equiv |{\vec k}|$.
This $k^4$-term implies that, depending on the sign of the coefficient
in front of the $k^4$-term, either the universe is very unstable on
small scales or the short wavelength fluctuations propagate with speed
larger than the speed of light. The background
classical-instabilities, due to the presence of a negative $k^4$-term,
must be forbidden, whereas there still exists controversy as to if the
superluminal propagation is compatible or not with experimental
observations
\cite{Bonvin:2006vc,Hashimoto:2000ys,Bruneton:2006gf,Ellis:2007ic,Babichev:2007dw}. Anyhow,
the scalar cosmological perturbation for the general MGM exhibits
highly non-trivial and interesting features that can be used to set
constraints on the MGM or may even leave detectable imprints on cosmic
structures due to the $k^4$-term.

These results in \cite{DeFelice:2009ak} were found for the vacuum
only, i.e.\ no fields other than the gravitational field were included
in the Lagrangian. An obvious and important extension of this work is
to add some matter fields into the MGM action and to see how the
results we found in the vacuum case, are affected by the existence of
matter.  This is the main purpose of this paper.  As a first and
simple extension, although useful in early-times cosmology, in this
paper, we will consider a general MGM together with a single real
scalar field $\varphi$, whose action is given by
\begin{equation}
S=\frac{M_P^2}{16\pi} \int \de^4x\sqrt{-g}~f(R,\GB,\varphi)
-\frac{1}{2} \int \de^4x\sqrt{-g} ~w \partial^\mu \varphi \partial_\mu \varphi. \label{action0}
\end{equation}
If $w$ depends on $\varphi$, we can set $w$ to be a constant by a
suitable field redefinition.  Therefore, we will consider $w$ as a
constant from the beginning.  In this setup, when $w= 0$, the
Lagrangian reduces to the model in vacuum
(\ref{actionvacuum})\footnote{When $w = 0$, $\varphi$ becomes an
  auxiliary field.  From a variation with respect to $\varphi$, we
  obtain $\varphi$ as a function of $R$ and $\GB$,
  i.e. $\varphi=\varphi (R,\GB)$.  Removing $\varphi$ in the action by
  using this relation yields an action which only contains $R$ and
  $\GB$.  Therefore, obtained action belongs to
  (\ref{actionvacuum}).}.  We can even remove $w$, if $w>0$, by
redefining the scalar field as ${\tilde \varphi}=\sqrt{w}\,\varphi$.
However, we will leave $w$ as a free parameter to enable us to
consider those cases for which $w<0$.  We do not assume any particular
functional form for $f(R,\GB,\varphi)$.

Although (\ref{action0}) is the basic action we consider, for
convenience of the actual analysis, we mainly use a different action,
equivalent to (\ref{action0}), which is given by
\begin{align}
S&=\frac{M_P^2}{16\pi} \int \de^4x\sqrt{-g}~\bigg[ f(\lambda,\sigma,\varphi)+(R-\lambda) F(\lambda,\sigma,\varphi)+(\GB-\sigma) \xi(\lambda,\sigma,\varphi) \bigg]-\frac{1}{2} \int \de^4x\sqrt{-g}~ w ~ \partial^\mu \varphi \partial_\mu \varphi, \nonumber \\
&=\frac{M_P^2}{16\pi}\int \de^4x\sqrt{-g}~\bigg[ RF(\lambda,\sigma,\varphi)+\GB\xi(\lambda,\sigma,\varphi)-V(\lambda,\sigma,\varphi) \bigg]-\frac{1}{2} \int \de^4x\sqrt{-g} ~w ~ \partial^\mu \varphi \partial_\mu \varphi, \label{action1}
\end{align}
where $\lambda$ and $\sigma$ are auxiliary fields and
\begin{alignat}{4}
F(\lambda,\sigma,\varphi)& \equiv \frac{\partial f}{\partial \lambda},&&\xi(\lambda,\sigma,\varphi) \equiv \frac{\partial f}{\partial \sigma}, \label{Fandxi}\\
V(\lambda,\sigma,\varphi)&\equiv \lambda F(\lambda,\sigma,\varphi)+\sigma&&\xi(\lambda,\sigma,\varphi)-f(\lambda,\sigma,\varphi).
\end{alignat}
By the following way, we can verify that the action (\ref{action1}) is
equivalent to (\ref{action0}). By variating $S$ with respect to
$\lambda$ and $\sigma$, we have the equations for $\lambda$ and
$\sigma$, which are given by
\begin{alignat}{4}
(R-\lambda) F_\lambda&+(\GB-\sigma) F_\sigma&&=0, \label{lambda1} \\
(R-\lambda) F_\sigma&+(\GB-\sigma) \xi_\sigma&&=0, \label{sigma1}
\end{alignat}
where $F_\lambda=\partial F/\partial\lambda$ etc. If the combination
$F_\lambda \xi_\sigma-F_\sigma^2$ does not vanish, the two equations
are independent and $\lambda$ and $\sigma$ are given by
\begin{align}
\lambda&=R, \\
\sigma&=\GB.
\end{align}
Eliminating $\lambda$ and $\sigma$ in the original action by using
these results, we find that $S$ in (\ref{action1}) reduces to
(\ref{action0}): the equivalence of these two actions also holds for
the equations of motion. If the combination $F_\lambda
\xi_\sigma-F_\sigma^2$ vanishes, (\ref{lambda1}) and (\ref{sigma1})
are not independent from each other.  This does not mean we can not
eliminate $\lambda$ and $\sigma$ from the original action.  Because
(\ref{lambda1}) is the condition that the action remains the same
under the variation, all the possible pairs of $\lambda$ and $\sigma$
that satisfy (\ref{lambda1}) give the same action. We can put any
$(\lambda, \sigma)$ we want into $S$ as long as $(\lambda, \sigma)$
are the solutions of (\ref{lambda1}).  Obviously, $\lambda=R$ and
$\sigma=\GB$ are the solution and we can put them into (\ref{action1})
to find that it recovers the original action (\ref{action0}).

In the next section, we take the action approach to study the
cosmological scalar perturbations. That is, we will expand the action
(\ref{action1}) to second order in the perturbation and reveal new
propagation properties due to the inclusion of the scalar field.
After the classical analysis of the perturbations, we will also
consider quantization of the perturbations.

\section{Action}
\subsection{Background and Perturbation Variable}

We consider a spacetime which slightly deviates from a FLRW universe.
We write the metric as
\begin{equation}
\de s^2=-(1+2\alpha)\, \de t^2-2 a (t) \partial_i \beta\, \de t\, \de x^i+a^2(t) (\delta_{ij}+2\phi \delta_{ij}+2\partial_i \partial_j \gamma)\, \de x^i \,\de x^j, \label{per1}
\end{equation}
where $a(t)$ is the scale factor and $\alpha, \beta, \phi$ and
$\gamma$ represent the metric perturbations.  We only consider the
scalar type perturbation, as both the vector and tensor modes'
propagation will not be affected---at linear order---by the presence
of the scalar field, albeit the modified background dynamical
evolution. We also decompose the scalar field as $\varphi+\delta
\varphi$, where $\varphi$ is the background value and $\delta \varphi$
is the perturbation.  Since $f$ is a function of $R$ and $\GB$, which
must be expanded, $F$ and $\xi$ defined by (\ref{Fandxi}) also acquire
perturbations, which we denote as $\delta F$ and $\delta \xi$.  For
later convenience, we define $Z$ and $\chi$ by the following equation,
\begin{equation}
Z\equiv \frac{\partial f}{\partial \varphi}, \hspace{10mm} \chi \equiv a(\beta+a {\dot \gamma}).
\end{equation}

There are degrees of freedom of choosing $\Sigma_t$ which is a three dimensional hyper-surface of 
constant $t$. 
Changing from $\Sigma_t$ to ${\tilde \Sigma_t}$ corresponds to a time coordinate transformation: $t\to t+T(t,x^i)$. 
Under this transformation, the perturbation variables transform as
\begin{align}
{\tilde \alpha}&=\alpha-{\dot T}, \\
{\tilde \phi}&=\phi-H T, \\
{\tilde \chi}&=\chi-T, \label{gtchi} \\
{\tilde {\delta \varphi}}&=\delta \varphi-{\dot \varphi}T, \\
{\tilde {\delta F}}&=\delta F-{\dot F}T, \\
{\tilde {\delta \xi}}&=\delta \xi-{\dot \xi}T, \\
{\tilde {\delta Z}}&=\delta Z-{\dot Z}T.
\end{align}

\subsection{First order action}

To obtain the background equations of motion, we need to expand the
action (\ref{action1}) at first order in the perturbation variables.  We find
that, at first order, the action is given by
\begin{alignat}{4}
S^{(1)}&= \frac{M_p^2}{16 \pi}\int \de^3 x \de t a^3 \bigl[ &&3 \bigl( -V +6{\dot H}F+12 H^2 F+24 H^4 \xi+24 H^2 {\dot H}\xi \bigr)\phi +24 H \bigl( F+4H^2 \xi+2H{\dot H} \xi \bigr) {\dot \phi} \nonumber \\
&&&+6(F+4H^2 \xi )\, {\ddot \phi}-\bigl( V+6{\dot H} F+12 H^2 F+72 H^2 {\dot H} \xi+72 H^4 \xi \bigr)\, \alpha \nonumber \\
&&&-6H(F+4H^2 \xi ) {\dot \alpha}+Z \delta \varphi \bigr]+\int \de^3 x \de t a^3 \bigg[ \frac{w}{2} {\dot \varphi}^2 (\alpha+3\phi)+w {\dot \varphi} {\dot {\delta \varphi}} \biggr]. \label{firstorderaction}
\end{alignat}
The variation with respect to the auxiliary fields $\delta \lambda$
and $\delta \sigma$ can also contribute to $S^{(1)}$.  It gives the
background equations $\lambda=R$ and $\sigma=G$.  In
(\ref{firstorderaction}), we have already replaced $\lambda ,\sigma$
by $R, G$ respectively and eliminated terms proportional to $\delta
\lambda$ and $\delta \sigma$.

By integrating by parts, we find that the condition $S^{(1)}=0$
yields the background equations which are given by
\begin{align}
3H^2&=\frac{1}{F} \left(\frac12\, V-3H{\dot F}-12H^3 {\dot \xi} +\frac{4\pi w}{M_P^2} {\dot \varphi}^2 \right),\\
{\dot H}&=\frac{1}{2F+8H{\dot \xi}} \left[ -{\ddot F}+H{\dot F}-4H^2 ({\ddot \xi}-H{\dot \xi}) -\frac{8\pi w}{M_P^2 } {\dot \varphi}^2 \right],\\
w {\ddot \varphi}&+3w H {\dot \varphi}-\frac{M_P^2}{16\pi}Z=0.
\end{align}

\subsection{Second order action}

To obtain the perturbation equations, we need to expand the action up to
second order in the perturbation variables.  We denote the second order action
as $S^{(2)}$.  Although the calculation is straightforward, the
expressions at each step of the algebraic calculation become very
long.  Therefore we shall be content with explaining what we do at
each calculation step and giving the final expression of the second order
action which contains only dynamical fields.

The field variables we can perturb as free variables are
$\alpha$, $\chi$, $\phi$, $\delta \varphi$, $\delta F$ and $\delta
\xi$.  Note that $\delta \lambda$ and $\delta \sigma$ can be written
as linear combinations of $\delta F$, $\delta \xi$ and $\delta
\varphi$ by inverting the relations
\begin{alignat}{6}
\delta F&=F_\lambda \delta \lambda&&+F_\sigma \delta \sigma&&+F_\varphi \delta \varphi, \label{rela1}\\
\delta \xi&=F_\sigma \delta \lambda&&+\xi_\sigma \delta \sigma&&+\xi_\varphi \delta \varphi. \label{rela2}
\end{alignat}
Just for practical convenience, we will use $\delta F$ and $\delta \xi$
instead of $\delta \lambda$ and $\delta \sigma$.

By using those fields ($\alpha$, $\chi$, $\phi$, $\delta \varphi$,
$\delta F$, $\delta \xi$), we can expand the action up to second order
in those fields.  Next, we chose a suitable gauge to make the
calculation easier, the so-called Modified Gravity Models Gauge
(MGMG), where $\delta F +4H^2 \delta \xi=0$. This gauge condition
uniquely fixes the time slicing. By this gauge condition, we can
remove $\delta F$ from $S^{(2)}$, so that $S^{(2)}$ is a quadratic
functional of $\{ \alpha_M, \chi_M, \phi_M, \delta \varphi_M, \delta
\xi_M \}$, where the index $M$ indicates that we are working in the
MGMG. After a few integration by parts, we find that $\alpha_M,
\chi_M$ and $\delta \xi_M$ are auxiliary fields.  Eliminating these
fields by using their equations of motion, $S^{(2)}$ becomes a
functional of $\phi_M$ and $\delta \varphi_M$ only. Performing other integrations
by parts, we find that $S^{(2)}$ can be finally written as
\begin{alignat}{4}
S^{(2)}&=\frac{M_p^2}{16 \pi} \int \de t \de^3\vec x ~a^3(t) \bigl[ && A_{ab}(t){\dot V_a} {\dot V_b}
+a(t)^{-2} B(t)\epsilon_{ab} {\vec \nabla} {\dot V_a} \cdot {\vec \nabla} V_b
-a(t)^{-4} D_{ab}(t) \triangle V_a  \triangle V_b 
-a(t)^{-2}E_{ab}(t) {\vec \nabla} V_a \cdot {\vec \nabla} V_b \notag \\
&&& +C(t)\epsilon_{ab} {\dot V_a} V_b+M_{ab}(t) V_a V_b \bigr]. \label{secondaction}
\end{alignat}
Here $V_a$ with $(a=1,2)$ is defined as
\begin{equation}
V_1 = \phi_M, \qquad V_2=\delta \varphi_M.
\end{equation}
The matrices $A$, $D$, $E$ and $M$ are given by background
quantities. The matrix $\epsilon_{ab}$ is the pure antisymmetric matrix 
with element $\epsilon_{12}=1$. Furthermore, the only
non-zero matrix element for $D$ is $D_{11}$. In the short wavelength
regime, $A$, $B$, $D$ and $E$ alone determine the propagation nature of
the perturbations.  Explicit expressions of these matrices are given in
the appendix.

From this action, we can derive some important results regarding the
propagation properties of the short wavelength modes.  First, the
determinant of $A$ does not vanish in general.  Therefore, there are
two independent dynamical perturbation fields (hence four degrees of
freedom).  The reason why we get two dynamical fields is clear. In the
absence of the scalar field, i.e.\ in vacuum, it was found that
there is only one dynamical field coming from the metric perturbation.
Because we are now adding a dynamical scalar field which has nothing
to do with the gravitational sector, it is not surprising that this
scalar field brings new degrees of freedom.  As an obvious extension,
we will get $2(N+1)$ degrees of freedom for the scalar perturbations if
we add $N$ dynamical and independent scalar fields into the action.

In order not to have ghosts, we require that the two eigenvalues of $A_{ab}$ must be positive.
This condition is equivalent to the following conditions,
\begin{equation}
\det A > 0, \qquad\Tr A>0. \label{no-ghost}
\end{equation}

Second important result is that $D$ is not a zero matrix in general.
Therefore, the $k^4$-term which has already been present in the vacuum
case is still present even if the scalar field is added into the
theory. Explicit calculations show that only $D_{11}$ does not vanish.
Therefore, among the two dynamical fields, only one acquires
$k^4$-propagation. This, we think, is the most important contribution
of this paper. In fact, since $V_1$ represents the metric and $V_2$
describes the perturbation of the scalar field, we can say that only
the metric perturbation acquires the $k^4$-term behaviour.  We can
understand this result as follows.  Since the $k^4$-term is already
present in the vacuum case, the appearance of the $k^4$-term is solely
related to the nature of the modification of gravity.  As for the
scalar fields, it has a standard kinetic term and it is hard to
imagine that it would give an additional $k^4$-term contribution from
its kinetic term.  In fact, we find that $D_{11}$ is proportional to
the combination $F_\lambda \xi_\sigma-F_\sigma^2$ (see appendix),
which distinguishes classes of modifications of gravity.  Therefore,
$F_\lambda \xi_\sigma-F_\sigma^2=0$ is still applicable as a
sufficient condition for the absence of the $k^4$ propagation.  Below,
we will consider a general case where $F_\lambda \xi_\sigma-F_\sigma^2
\neq 0$, whereas, in the next section, we will consider some special
case where, for example, $F_\lambda \xi_\sigma-F_\sigma^2 = 0$.

From (\ref{secondaction}), we can derive the equations of motion for
$V_a$.  Using a Fourier transformation, the equations of motions for
the two modes are
\begin{equation}
\frac{{\bigl( a^3 A_{ab} {\dot V_b} \bigr)}^{\cdot}}{a^3}+\frac{k^2}{2a^5} {\left( a^3 B \epsilon_{ab} V_b \right)}^{\cdot}+\frac{k^2}{2a^2} B \epsilon_{ab} {\dot V_b}+D_{ab} \frac{k^4}{a^4} V_b+E_{ab} \frac{k^2}{a^2} V_b+\frac{{\left( a^3 C \epsilon_{ab}V_b \right)}^{\cdot}}{2a^3}+\frac{1}{2} C\epsilon_{ab} {\dot V_b}-M_{ab} V_b=0.
\end{equation}
In the short wavelength limit ($k$ large), this vectorial equation can
be approximated as
\begin{equation}
A_{ab} {\ddot V_b}+\frac{k^2}{a^2} B \epsilon_{ab} {\dot V_b}+D_{ab} \frac{k^4}{a^4} V_b+E_{ab} \frac{k^2}{a^2} V_b=0.
\end{equation}
From this equation, we can derive a dispersion relation. By denoting
the angular frequency for the mode $k$ as $\omega$, the dispersion
relation is given by
\begin{equation}
\det\!\left( -\omega^2 A+i\omega \frac{k^2}{a^2} B\epsilon+\frac{k^4}{a^4} D+\frac{k^2}{a^2} E \right)=0.  \label{dispersion}
\end{equation}
This is a quadratic equation in $\omega^2$ and it has two solutions.
As we discussed before, only two modes among four acquire $k^4$
propagation.  From (\ref{dispersion}), to leading order in $k$, we
find that the angular frequency for those modes can be written as
\begin{equation}
\omega^2 \approx \frac{B^2+\Tr A \,\Tr D-\Tr (AD)}{\det A} \frac{k^4}{a^4}=\frac{B^2+A_{22}D_{11}}{A_{11}A_{22}-A_{12}^2} \frac{k^4}{a^4}.
\end{equation}
Using the explicit expression for the matrix elements given in the appendix,
$\omega^2$ is given by
\begin{equation}
\omega^2=-\frac{64}{3} \frac{{\dot H}^2 (F_\lambda \xi_\sigma-F_\sigma^2)}{(F+4H{\dot \xi}) (F_\lambda+8H^2 F_\sigma+16H^4 \xi_\sigma)}\frac{k^4}{a^4}. \label{dispersion2}
\end{equation}
It is impressive that this expression, in form, is exactly the same as
that in the vacuum case \cite{DeFelice:2009ak}.  However, by using (\ref{rela1}), we can
replace, for example, $F_\lambda$ with different background quantities
such as ${\dot F}, F_\sigma, F_\varphi$ and ${\dot \varphi}$.  After
this replacement, $\omega$ explicitly depends on the scalar fields and
is not anymore the same as the vacuum case.  There is no clear
distinction between implicit and explicit dependence of the scalar
field on $\omega^2$. To conclude, the introduction of the scalar field
does not remove the $k^4$ propagation but changes the background
dependent coefficients in the dispersion relation.

As discussed in detail in \cite{DeFelice:2009ak}, if the right hand
side of (\ref{dispersion2}) is negative, the perturbation grows
exponentially in time. The growth rate is higher for shorter
wavelength (proportional to $k^2$).  Therefore, the FLRW universe is
very unstable.  If the RHS is positive, the perturbation propagate as
a wave.  In this case, the propagation speed (group velocity) is
linearly dependent on $k$.  Therefore, the propagation speed is in
general superluminal at small scales.

As for the dispersion relation for the other two modes,
we find that $\omega$, to leading order in $k$, can written as
\begin{equation}
\omega^2= \frac{D_{11} E_{22}}{B^2+A_{22} D_{11}} \frac{k^2}{a^2}=\frac{k^2}{a^2}.
\end{equation}
Therefore, these modes propagate at the speed of light.

\section{Special cases}

In the previous section, we have found that the scalar perturbations
have four degrees of freedom (two dynamical fields).  Among them, two
modes acquire a $k^4$ propagation which depends on background
quantities, whereas the other two propagate at the speed of light.
However, there are special cases for which we can not apply this
generic analysis. This is the case if $f(R,\GB,\varphi)$ satisfies the
special relation $F_\lambda \xi_\sigma-F_\sigma^2=0$, or $w=0$, or the
universe undergoes de-Sitter expansion.  In the following, we will
consider each case separately.

\subsection{Case I: $F_\lambda \xi_\sigma-F_\sigma^2=0$, $w\neq0$}

In this case, both $B$ and $D_{ab}$ vanish identically.  Therefore,
the dispersion relation (\ref{dispersion}) simplifies to
\begin{equation}
\det \left( -\omega^2 A+\frac{k^2}{a^2} E \right)=0.  \label{dispersionsp}
\end{equation}
Explicit forms of the matrix elements of $A$ and $E$ are given in the appendix.
This equation yields a quartic equation for $\omega$, which is given by
\begin{equation}
\det A ~\omega^4-\frac{k^2}{a^2} \left[ \Tr A\, \Tr E-\Tr  (AE) \right] \omega^2+\frac{k^4}{a^4}\det E=0. \label{sp1}
\end{equation}
This equation always has real solutions for $\omega^2$. Explicit forms
of $\det A,~\det E$ and $\Tr A\, \Tr E-\Tr (AE)$ in this case are also
given in the appendix.

Eq.~(\ref{sp1}) has two different solutions for $\omega^2$, in
general. Interestingly $\omega^2=k^2/a^2$ is not
generally a solution of (\ref{sp1}) any longer. This can be seen by
substituting this ansatz into (\ref{sp1}).  Under this substitution,
the LHS of (\ref{sp1}) becomes
\begin{equation}
\frac{k^4}{a^4}-\frac{k^4}{a^4}\frac{\Tr A \Tr E-\Tr  (AE)}{\det A}+\frac{k^4}{a^4}\frac{\det E}{\det A}=-\frac{4}{3} \frac{ M_P^2 {\dot H}^2 {( {\dot F} \xi_\sigma-{\dot \xi} F_\sigma )}^2}{ \pi w {\dot \varphi}^2 (F+4H{\dot \xi}) {(F_\sigma+4H^2 \xi_\sigma)}^2} \frac{k^4}{a^4}. \label{dissol}
\end{equation}
Assuming the universe is not de Sitter (the de Sitter case will be
studied as another special case), this does not vanish unless ${\dot F} \xi_\sigma-{\dot \xi} F_\sigma=0$. 
This happens, for example, when $f=f(R,\GB)-U(\varphi)$.

If at least one of the solutions for $\omega^2$ is negative, the
perturbations are classically unstable.  The condition for not having ghosts modes (\ref{no-ghost}), and the condition that the two $\omega^2$ are both positive are equivalent to
\begin{equation}
\Tr A>0,\qquad\det A >0,\qquad\det E >0,\qquad\Tr A \Tr E-\Tr (A\,E)>0.
\end{equation}

Just for the purpose of seeing how our findings are applied to some
concrete models, let us consider simple examples which have been
frequently studied in literature.  The first one is
\cite{Starobinsky:1980te,Capozziello:2002rd,Carroll:2003wy}
\begin{equation}
f(R,\GB,\varphi)=f(R,\varphi).
\end{equation}
In this case, we have ${\dot \xi}=\xi_\sigma=F_\sigma=0$ and
\begin{equation}
\det A=\frac{3 \rew F {\dot F}^2}{ {({\dot F}+2HF)}^2},~~~\Tr A \Tr E-\Tr (A\,E)=2 \det A,~~~\det E=\det A.
\end{equation}
Therefore, the dispersion relation becomes
\begin{equation}
\det A \left( \omega^2-\frac{k^2}{a^2} \right)^2=0.
\end{equation}
Therefore, assuming $\det A \neq 0$, all the four modes propagate at
the speed of light.  There are, however, well-known exceptions, i.e.\
GR and the Brans-Dicke theory.  In GR, $F=1$ and ${\dot F}=0$ and
hence $\det A=0$. This means that one of the fields does not
propagate.  The perturbation fields in (\ref{secondaction}) are
defined on MGMG. In GR, this gauge condition is automatically
satisfied, so that there still remains some gauge degrees of freedom.
By using this remaining gauge, we can always set one of the two
perturbation fields to be zero.

For the Brans-Dicke theory, $f(R,\GB,\varphi)$ is given by
\begin{equation}
f(R,\GB,\varphi)=\frac{2\pi}{M_P^2}\varphi^2 R.
\end{equation}
Using a new field defined by $\psi=\varphi^2/8$, the action indeed
reduces to the original one proposed in \cite{Brans:1961sx},
\begin{equation}
S=\int \de^4 x \sqrt{-g} \left( \psi R-\frac{w}{\psi}\partial^\mu \psi \partial_\mu \psi \right).
\end{equation}

Therefore, for Brans-Dicke theory, $F= 2\pi \varphi^2/M_P^2$ and
$\xi=0$. Then MGMG imposes that $V_2=\delta \varphi_M=0$.  Therefore,
only the $V_1$ field is dynamical and $A_{11}$ and $E_{11}$ determine alone the
propagation nature of the perturbations.  Applying the explicit
expressions for $A_{11}$ given in the appendix, we find that $A_{11}$
is given by
\begin{equation}
A_{11}=(2w+3) \frac{\varphi^2 {\dot \varphi}^2}{ 4{( {\dot \varphi}+H\varphi)}^2 }.
\end{equation}
Therefore, a ghost appears when $w < -3/2$, which agrees with well
known results \cite{Dicke:1961gz}.  Furthermore, it can be shown that the
propagation speed of the perturbations is the speed of light.

Another similar case is the Lagrangian with
\begin{equation}
  \label{eq:spsp}
  f(R,\GB,\varphi)= F(\varphi)R+\xi(\varphi)\GB-U(\varphi)\, ,
\qquad{\rm and}\qquad w\neq0.
\end{equation}
In this case the number of propagating fields is only one, as the MGMG
corresponds to setting $\delta\varphi=0$. This case was the only one
studied in \cite{Hwang:2005hb}, although the authors claimed to
have studied $ f(R,\GB,\varphi)=
f_1(\varphi,R)+\xi(\varphi)\GB-U(\varphi)$, with general $f_1$, which has, instead, two
propagating fields.

A second example for a special case is given by \cite{DeFelice:2008wz,DeFelice:2009aj},
\begin{equation}
f(R,\GB,\varphi)=R+f_1(\GB,\varphi).
\end{equation}
In this case, we have $F=1,~{\dot F}=0$ and
\begin{align}
\det A &=\frac{12 H^2 {\dot \xi}^2 \rew (1+4H{\dot \xi})}{ {(1+6H{\dot \xi})}^2 }, \\
\Tr A \Tr E-\Tr (AE)&=\frac{8 \rew {\dot \xi}^2}{ {(1+6H {\dot \xi})}^2 }(3H^2+14 H^3 {\dot \xi}+2{\dot H}-2H^2 {\ddot \xi}+8H{\dot H}{\dot \xi}), \\
\det E&=\rew \frac{4 {\dot \xi}^2}{ {(1+6H{\dot \xi})}^2} \left( 16H(H^2+{\dot H}){\dot \xi}-4H^2 {\ddot \xi}+3H^2+4{\dot H} \right).
\end{align}
Then the dispersion relation becomes
\begin{equation}
\left( \omega^2-\frac{k^2}{a^2} \right) \left[ \omega^2-\left( 1+\frac{2 {\dot H}}{H^2}+\frac{\rew}{(1+4H {\dot \xi}) H^2} {\dot \varphi}^2 \right)\frac{k^2}{a^2} \right]=0.
\end{equation}
We find that two out of four modes propagate at the speed of light.
This result is consistent with the discussion we did just after
(\ref{dissol}). As for the other two modes, the square of the
propagation speed is
\begin{equation}
c_s^2=1+\frac{2 {\dot H}}{H^2}+\frac{16\pi w}{M_P^2(1+4H {\dot \xi}) H^2} {\dot \varphi}^2. \label{scalarcs2}
\end{equation}
In vacuum, we would get $1+2{\dot H}/H^2$ which agrees with
\cite{DeFelice:2009ak}. There, it was found that the universe must be
accelerating in order to avoid a negative $c_s^2$, furthermore the
propagation was becoming superluminal during super-acceleration,
i.e.\ ${\dot H}>0$.  In the presence of a scalar field,
(\ref{scalarcs2}) implies that, depending on the sign of $w\,(1+4H{\dot
  \xi})$, the scalar field contribution can shift $c_s^2$ toward either
positive or negative values.  Therefore, in principle, it is possible
to have positive and sub-luminal $c_s^2$ even in a decelerating
universe.

\subsection{Case II: non-propagating field---$w=0$}

Let us here discuss the case in which $w=0$. This is a special
case, which can be understood by noticing that in this case $\varphi$
becomes an auxiliary field. In fact, after integrating it out, one can
show that the Lagrangian reduces to the vacuum $f(R,\GB)$ case already discussed in
\cite{DeFelice:2009ak}. However it is also possible to keep the field
$\varphi$, and expand the action at second order in all the fields. In
this case, choosing the MGMG gauge, one finds only one propagating
field, $\phi_M$, as expected. However, the $k^4$-term is still present, and one
finds that for large $k$, the angular frequency is given by
\begin{equation}
  \label{eq:om0SP1}
  \omega^2=-\frac{64}3\,\frac{(\xi_\sigma F_\lambda Z_\varphi+
    2F_\sigma F_\varphi \xi_\varphi-Z_\varphi F_\sigma^2-
    F_\lambda \xi_\varphi^2-\xi_\sigma F_\varphi^2)\, \dot H^2}{(F+4H^2\dot\xi)[(F_\lambda Z_\varphi-F_\varphi^2)+
    8H^2(F_\sigma Z_\varphi-F_\varphi \xi_\varphi)+16H^4(\xi_\sigma Z_\varphi-\xi_\varphi^2)]}\,\frac{k^4}{a^4}\, .
\end{equation}
This case is, in a sense, discontinuous from the general case, as the
degrees of freedom reduce by two, since the field $\delta \varphi_M$ becomes
itself dependent on the field $\phi_M$. Eq.\ (\ref{eq:om0SP1}) implies
that the $k^4$-term will be vanishing, for example, if $F=F(\varphi)$
and $\xi=\xi(\varphi)$.

\subsection{Case III: de Sitter---${\dot H}=0$}

In this case, since both $\delta F$ and $\delta \xi$ are gauge
invariant, we can not impose MGMG in general and (\ref{secondaction})
is not applicable anymore. Instead of using the variables involved in
the MGMG, we find it convenient to use gauge invariant variables
$\Phi$ defined by
\begin{equation}
\Phi=-\frac{\delta F+4H^2 \delta \xi}{2F_0},
\end{equation}
and $\delta \varphi$.  Note that in de Sitter all background
quantities except for the scale factor are constants.  Using these
perturbation variables, we find that the second order action can be
written as
\begin{alignat}{4}
S^{(2)}&=\frac{M_P^2}{16\pi} \int \de^3x \de t~a^3 \Bigl[&& 6F \bigl( {\dot \Phi}^2-a^{-2}{\vec \nabla} \Phi \cdot {\vec \nabla} \Phi \bigr)+\frac{8\pi w}{M_P^2} \bigl( \dot {\delta \varphi}^2-a^{-2} {\vec \nabla} \delta \varphi \cdot {\vec \nabla} \delta \varphi \bigr)\notag\\
&&&{}-m_1^2 \Phi^2-m_2^2 \Phi \delta \varphi-m_3^2 ({\delta \varphi)}^2 \Bigr],
\end{alignat}
where $m_1^2,m_2^2$ and $m_3^2$ are given by
\begin{align}
m_1^2&=\frac{2F (F-192 H^6 \xi_\sigma-96 H^4 F_\sigma-12H^2 F_\lambda)}{F_\lambda+8H^2 F_\sigma+16H^4 \xi_\sigma}, \\
m_2^2&=\frac{2F (F_\varphi+4H^2 \xi_\varphi)}{F_\lambda+8H^2 F_\sigma+16H^4 \xi_\sigma}, \\
m_3^2&=\frac{16H^4 \xi_\varphi^2-F_\lambda Z_\varphi+F_\varphi^2-16 H^4 \xi_\sigma Z_\varphi+8H^2 F_\varphi \xi_\varphi-8H^2 Z_\varphi F_\sigma}{2(F_\lambda+8H^2 F_\sigma+16H^4 \xi_\sigma)}.
\end{align}
We find that both fields propagate at the speed of light.  In order
not to have ghosts, the conditions $F>0$ and $w>0$ are required.

\section{Quantization}
Up to here, we have considered only the classical theory of the
perturbations.  Although we have found non-trivial and interesting
results at the classical level, nothing prevents us from going to
quantize the perturbations.  An obvious important application of the
quantization will be the quantum generation of the curvature
perturbations during inflation.  In this case, inflation may be caused
by the modification of gravity or by the scalar field $\varphi$
(inflaton). Then, the observed curvature perturbations will be a
mixture of the scalar metric perturbations and the scalar field
perturbations.  Another possible application would be preheating
\cite{Traschen:1990sw, Kofman:1994rk} after inflation, where the
oscillations of $\varphi$ (if it is an inflaton) induces rapid
creation of both $V_1$ and $V_2$ quanta.  These applications are
interesting by themselves and will be discussed elsewhere.  In this
section, we will provide general procedure to quantize the
perturbations.

Our starting point is the second order action for the perturbations,
which is given by Eq.~(\ref{secondaction}). To quantize it, it is
convenient to write the action in terms of canonical fields.  Assuming
$\det A \neq 0$, there always exists a time-dependent matrix
$Z_{ab}(t)$ such that
\begin{equation}
a^3(t)\, Z^T A Z=I,
\end{equation}
where $I$ is the identity matrix.  Then the fields defined by $\psi_a
\equiv {\left( Z^{-1} \right)}_{ab} V_b$ are the canonical fields, as
their kinetic term becomes
\begin{equation}
S^{(2)} = \int \de t \de ^3x\, \frac{1}{2} \delta_{ab} {\dot \psi}_a {\dot \psi}_b+\cdots.
\end{equation}
However, as $Z$ depends on the time, we also get time derivative of
$Z$ contributions coming from the first two terms in
(\ref{secondaction}).  This yields additional terms each of which has
the same form as one of the last three terms in (\ref{secondaction}),
which can be absorbed into a redefinition of the background dependent
matrices. Therefore, without loss of generality, we can write the
second order action like
\begin{alignat}{4}
S^{(2)}&=\int \de t\, \de ^3x~{\cal L}&& \notag \\
&=\int \de t\, \de^3x\, \biggl[ &&\frac{1}{2} \delta_{ab}{\dot \psi_a} {\dot \psi_b}  + b(t) \epsilon_{ab} {\vec \nabla} {\dot \psi_a} \cdot {\vec \nabla} \psi_b-\frac{1}{2} d_{ab}(t) \triangle \psi_a  \triangle \psi_b -\frac{1}{2} H_{ab}(t) {\vec \nabla} \psi_a \cdot {\vec \nabla} \psi_b \nonumber \\
&&&+c(t) \epsilon_{ab} {\dot \psi_a} \psi_b-\frac{1}{2} m_{ab}(t) \psi_a \psi_b \biggr], \label{canonicalaction}
\end{alignat}
where the background dependent matrices for each term are related to
the matrices in (\ref{secondaction}) by $Z$.  Although we can further
perform an orthogonal transformation in order to make $d_{ab}(t)$
diagonal---still keeping the kinetic term diagonal---we leave it as a
general symmetric matrix.

We can define canonical momenta conjugate to $\psi_a$ as
\begin{equation}
\pi^a = \frac{\partial {\cal L}}{\partial {\dot \psi_a}}={\dot \psi_a}-b\epsilon_{ab} \triangle \psi_b+c \epsilon_{ab}\psi_b.
\end{equation}
We impose the canonical quantization conditions, which are given by
\begin{align}
&[ {\hat \psi}_a (t,{\vec x}),{\hat \psi}_b (t,{\vec y})]=[ {\hat \pi}^a (t,{\vec x}),{\hat \pi}^b (t,{\vec y})]=0, \label{ccm1} \\
&[ {\hat \psi}_a (t,{\vec x}),{\hat \pi}^b (t,{\vec y})]=i \delta_a^b \delta ({\vec x}-{\vec y}). \label{ccm2}
\end{align}
The Heisenberg equations of motion, with the help of the above commutation
relations, yield the following evolution equations for ${\hat \psi_a}$,
\begin{equation}
{\ddot {\hat \psi}_a}-2 b \epsilon_{ab} \triangle {\dot {\hat \psi}_b}+2 c\epsilon_{ab} {\dot {\hat \psi}_b}+ d_{ab} \triangle^2 {\hat \psi}_b-( H_{ab}-{\dot b} \epsilon_{ab}) \triangle {\hat \psi}_b+(m_{ab}+{\dot c} \epsilon_{ab}) {\hat \psi}_b=0. \label{operatorevolve}
\end{equation}
According to \cite{GrootNibbelink:2001qt}, it is always possible to
write solutions of the equations consistent with the commutation
relations as
\begin{equation}
{\hat \psi}_a (t,{\vec x})= \int \frac{d^3 k}{ {(2\pi)}^{3/2} } \left[ U_{ab} (t,{\vec k}) {\hat a}_{b} ({\vec k}) e^{i {\vec k} \cdot {\vec x}}+ U_{ab}^\ast (t,{\vec k}) {\hat a}^\dagger_{b} ({\vec k}) e^{-i {\vec k} \cdot {\vec x}} \right].
\end{equation}
Here, the time-independent operators ${\hat a}_a({\vec k})$ satisfy the following commutation relations,
\begin{alignat}{4}
&[{\hat a}_a({\vec k_1}),& {\hat a}_b({\vec k_2})]&=[{\hat a}_a^\dagger ({\vec k_1}), {\hat a}_b^\dagger ({\vec k_2})]=0, \label{com1} \\
&[{\hat a}_a ({\vec k_1}),& {\hat a}_b^\dagger ({\vec k_2})]&=\delta_{ab} \delta ({\vec k_1}-{\vec k_2}). \label{com2}
\end{alignat}
The matrix $U_{ab} (t,{\vec k})$ is a collection of c-number time-dependent functions
which satisfy both the classical equation of motion,
\begin{equation}
{\ddot U_{ab}}+2 (b k^2+ c) \epsilon_{ac} {\dot U_{cb}}+ d_{ac} k^4 U_{cb}+( H_{ac}-{\dot b} \epsilon_{ac}) k^2 U_{cb}+(m_{ac}+{\dot c} \epsilon_{ac}) U_{cb}=0, \label{evolU}
\end{equation}
and the constraints given by
\begin{align}
&U U^{\ast T}-U^\ast U^T=0, \\
&{\dot U} {\dot U}^{\ast T}-{\dot U}^\ast {\dot U}^T=-2i (bk^2+c)\epsilon, \\
&U {\dot U}^{\ast T}-U^\ast {\dot U}^T=i I.
\end{align}
These constraints are necessary for the consistency of the commutation
relations (\ref{com1}), (\ref{com2}) with the canonical commutation
relations (\ref{ccm1}), (\ref{ccm2}).  By using the evolution
equations (\ref{evolU}), we can check that these constraints are
satisfied at any time once they are imposed at one time.  A simple
example of $U$ and ${\dot U}$ that satisfy the constraints, is given
by $U=I, {\dot U}=-\frac{i}{2} I-(bk^2+c)\epsilon$ at some initial
time. However, there are infinite other combinations that satisfy the
same constraints. Which $U$ we should use will depend on the physical
situation (initial conditions) we are interested in. Here we do not
specify the form of $U$.

As is well-known, there exists a state $|0 \rangle$ which is defined
by
\begin{equation}
{\hat a}_a ({\vec k}) |0 \rangle = 0.
\end{equation}
Then we can construct a particle-basis by applying the creation
operators ${\hat a}^\dagger_a({\vec k})$ to the vacuum state.  Once
this is done, we can calculate expectation values of any operator for
the state constructed in this way.  For example, the ``vacuum''
two-point function is given by
\begin{equation}
\langle 0| {\hat \psi}_a (t,{\vec x}) {\hat \psi}_b (t,{\vec y}) | 0 \rangle = \int \frac{\de^3 k}{ {(2\pi)}^3 } U_{ac} (t,{\vec k}) U^\ast_{bc} (t,{\vec k})e^{i {\vec k} \cdot ({\vec x}-{\vec y})}.
\end{equation}

Before closing this section, let us finally make a comment.  The
evolution equations (\ref{operatorevolve}) in Fourier space can be
seen as the equations of motion for an anisotropic charged harmonic
oscillator with time-dependent spring force in a two-dimensional
plane, in the presence of a time-dependent magnetic field. In fact,
$\psi_1$ and $\psi_2$ represent the $x$ and $y$ coordinate of the
point mass respectively, and the magnetic field whose strength is
$2(bk^2+c)$ (for unit charge) is parallel to the $z$-axis.  Therefore,
the quantization of (\ref{canonicalaction}) is equivalent to the
quantum mechanics of the harmonic oscillator under a time-dependent
magnetic field.  For time-independent spring forces in the presence of
a time-independent magnetic field, \cite{Rebane:1969,Ternov:1970} gave
the analytic wave function and energy spectrum, which, in terms of the
Heisenberg picture we consider in this paper, means that analytic
solutions of (\ref{evolU}) were found.  For the time-dependent case,
\cite{Abdalla:1988a, Abdalla:1988b} found analytic wave function for an
isotropic harmonic oscillator.  For the general case, finding the
analytic solution of (\ref{evolU}) is not an easy task and it is
beyond the analysis presented in this paper.

\section{conclusion}

We have studied the scalar type cosmological perturbations for modified
gravity models with a single scalar field.  Followings are our
findings.

\begin{itemize}
\item There are two dynamical perturbation variables, one from the
  gravity sector and one from the scalar field sector.  As a result,
  we have four degrees of freedom.  However, there are a few exceptions.  
  In GR, as is well known, the scalar mode of gravity does
  not propagate and we have only one dynamical perturbation
  variable. Also the case $w=0$---no kinetic term for the scalar
  field---leads to only one propagating degree of freedom, which
  arises because gravity has been modified.

\item Even in the presence of the scalar field, the action, expanded
  at second order in the fields, contains a term proportional to
  $k^4$. Therefore, inclusion of the matter field (at least in the
  form of a single scalar field) does not eliminate such term.  To be
  more precise, among the four independent modes, two modes acquire a
  $k^4$ propagation.  The propagation speed of these modes is affected
  by the presence of the scalar field through the modification of the
  background dynamics due to the scalar field.  Depending on the
  modified gravity model and background solution, propagation becomes
  superluminal or the universe is unstable on small scales.  The
  propagation speed of other two modes is equal to the speed of light.

\item If the model satisfies the special condition
  $F_\lambda\xi_\sigma-F_\sigma^2=0$ and $w\neq0$, the $k^4$-term
  disappears from the action.

\item For the special case $F_\lambda \xi_\sigma-F_\sigma^2=0$, the
  propagation of the short wavelength modes are dominated by the
  $k^2$-term.  If the combination ${\dot F}\xi_\sigma-{\dot
    \xi}F_\sigma$ does not vanish, none of the four modes propagates at
  the speed of light.  It is very interesting and highly non-trivial
  that two modes which propagate at the speed of light in the general
  case no longer propagate at the speed of light in the special case.

\item If the universe is de Sitter, irrespective of the MGM, all the
  modes propagate at the speed of light.  Both $F$ and $w$ must
  be positive to avoid ghosts.
\end{itemize}

\begin{figure}[ht]
\begin{center}
\includegraphics[width=10cm,keepaspectratio]{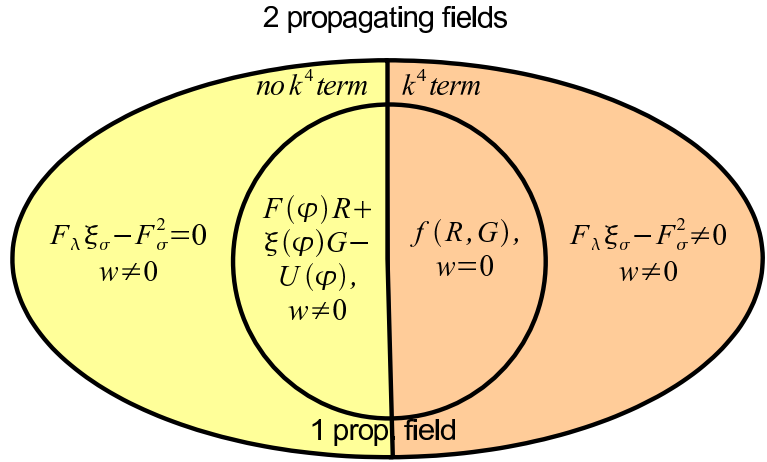}
\caption{Classification of the MGM according to the properties of the
  propagation for the perturbation fields. The models in the inner
  circle contain only one propagating field.  The models in the orange
  (darker) region have modes which have the $k^4$-propagation, whereas
  those in the yellow (lighter) region do not have such modes.}
\label{fig}
\end{center}
\end{figure}

In Fig.~\ref{fig} we summarize the classification of the MGM according to the properties of the propagation for the perturbation fields.

\begin{acknowledgments}
  We thank Jan Govaerts, Sean Murray, and Sebawe Abdalla for helpful
  discussions. This work is supported by the Belgian Federal Office
  for Scientific, Technical and Cultural Affairs through the
  Interuniversity Attraction Pole P6/11.
\end{acknowledgments}

\appendix

\section{Matrix elements  $A_{ab}$,  $D_{ab}$, and $B$ for the general case}
The matrix elements of $A_{ab}$ are given by
\begin{equation}
A_{11}=\frac{2 a_{11}}{J}, ~~~~~A_{12}=\frac{a_{12}}{J}, ~~~~~~A_{22}=\frac{8\pi}{M_P^2} w \frac{a_{22}}{J},
\end{equation}
where
\begin{align}
a_{11}&=(F+ 4\,H\dot\xi)\,\biggl( 768\,\xi_\sigma\,{H}^{8}{\dot\xi}^{2}+384\,F_\sigma\,{H}^{6}{\dot\xi}^{2}+384\,\xi_\sigma\,{H}^{6
}\dot F\,\dot\xi+64\,\xi_\sigma\,{\dot\varphi}^{2}\rew\,\dot\xi
\,{H}^{5}+48\,F_\lambda\,{\dot\xi}^{2}{H}^{4}\notag\\
&+192\,F_\sigma\,{H}^{4}\dot F\,\dot\xi+16\,{H}^{4}F\xi_\sigma\,{\dot\varphi}^{2}\rew+48\,\xi_\sigma\,{H}^{4}{\dot F}^{2}+32\,F_\sigma\,{\dot\varphi}^{2}\rew\,
\dot\xi\,{H}^{3}+24\,F_\lambda\,\dot F\,\dot\xi\,{H}^{2}\notag\\
&+8\,{H}^
{2}FF_\sigma\,{\dot\varphi}^{2}\rew+192\,{F_\sigma}^{2}\rew\,{\dot\varphi}^{2}{H}^{2}{\dot H}^{2}-192\,\rew\,{\dot\varphi}^{2}\xi_\sigma
\,F_\lambda\,{H}^{2}{\dot H}^{2}+24\,F_\sigma\,{H}^{2}{\dot F}^{2
}\notag\\
&+4\,F_\lambda\,{\dot\varphi}^{2}\rew\,\dot\xi\,H+3\,F_\lambda\,{\dot F}^{2}+FF_\lambda\,{\dot\varphi}^{2}\rew\biggr),\\
a_{12}&=\dot\varphi\,\rew\, (F+ 4\,H\dot\xi )  \,( 192\,\dot\xi\,\xi_\sigma\,{H}^{6}+96\,\dot\xi\,F_\sigma\,{H}^{4}+12\,\dot\xi\,F_\lambda\,{H}^{2}+32\,\xi_\sigma\,F{H}^{5}+16\,\xi_\sigma\,\dot F
\,{H}^{4}\notag\\
&+384\,{H}^{3}{F_\sigma}^{2}{\dot H}^{2}+16\,{H}^{3}F_\sigma\,F-384\,{H}^{3}\xi_\sigma\,{\dot H}^{2}F_\lambda+8\,F_\sigma\,\dot F\,{H}^{2}+2\,F_\lambda\,FH+\dot F\,F_\lambda),\\
a_{22}&=768 H^7 F {\dot \xi} \xi_\sigma+32 H^3 F {\dot F} F_\sigma+384 H^5 F {\dot \xi} F_\sigma+32 H^4 F^2 F_\sigma+64 H^5 F {\dot F} \xi_\sigma+192 H^4 {\dot F} {\dot \xi} F_\sigma+1152 H^6 {\dot \xi}^2 F_\sigma \nonumber \\
&+384 H^6 {\dot F} {\dot \xi} \xi_\sigma-3072 H^5 {\dot H}^2 {\dot \xi} (F_\lambda \xi_\sigma-F_\sigma^2)-768 H^4 {\dot H}^2 F (F_\lambda \xi_\sigma-F_\sigma^2)+{\dot F}^2 F_\lambda+64 H^6 F^2 \xi_\sigma+8 H^2 {\dot F}^2 F_\sigma \nonumber \\
&+4H^2 F^2 F_\lambda+4 HF {\dot F} F_\lambda+144 H^4 {\dot \xi}^2 F_\lambda+16 H^4 {\dot F}^2 \xi_\sigma+2304 H^8 {\dot \xi}^2 \xi_\sigma+48 H^3 F{\dot \xi} F_\lambda+24 H^2 {\dot F} {\dot \xi} F_\lambda,
\end{align}
where $J$ is defined by
\begin{align}
J &= 768 H^7 F {\dot \xi} \xi_\sigma+32 H^3 F {\dot F} F_\sigma+384 H^5 F {\dot \xi} F_\sigma+32 H^4 F^2 F_\sigma+64 H^5 F {\dot F} \xi_\sigma+192 H^4 {\dot F} {\dot \xi} F_\sigma+1152 H^6 {\dot \xi}^2 F_\sigma \nonumber \\
&+384 H^6 {\dot F} {\dot \xi} \xi_\sigma-3072 H^5 {\dot H}^2 {\dot \xi}( F_\lambda \xi_\sigma-F_\sigma^2)-64 H^2 {\dot H}^2 \rew {\dot \varphi}^2 (F_\lambda \xi_\sigma -F_\sigma^2)-768 H^4 {\dot H}^2 F (F_\lambda \xi_\sigma-F_\sigma^2)+{\dot F}^2 F_\lambda \nonumber \\
&+64 H^6 F^2 \xi_\sigma+8 H^2 {\dot F}^2 F_\sigma+4 H^2 F^2 F_\lambda+4HF {\dot F} F_\lambda+144 H^4 {\dot \xi}^2 F_\lambda+16 H^4 {\dot F}^2 \xi_\sigma+2304 H^8 {\dot \xi}^2 \xi_\sigma+48 H^3 F {\dot \xi} F_\lambda \nonumber \\
&+24 H^2 {\dot F} {\dot \xi} F_\lambda.
\end{align}
Because $A_{ab}$ is symmetric matrix, $A_{21}=A_{12}$.

The quantity $B$ is given by
\begin{equation}
B=-64 H{\dot H}^2 \rew {\dot \varphi}^2 ({\dot F}+4H^2 {\dot \xi}) \frac{(F_\lambda \xi_\sigma-F_\sigma^2)}{J}.
\end{equation}

The matrix element of $D_{11}$ is given by
\begin{equation}
D_{11}=-128 a {\dot H}^2 \frac{{({\dot F}+4H^2 {\dot \xi} )}^2 (F_\lambda \xi_\sigma-F_\sigma^2)}{J},
\end{equation}
and all the other components, as already said, are identically equal to zero.

Except for $E_{22}$, $E_{ab}$ appears in the dispersion relation at
the leading order only for the special case $F_\lambda
\xi_\sigma-F_\sigma^2=0$.  Therefore, we give their explicit
expressions for the special case.  Irrespective of modified gravity
models, $E_{22}$ is given by (\ref{h22}).
\begin{align}
E_{11}&=\frac{2}{ a {( {\dot F}+2HF+12H^2 {\dot \xi} )}^2} \bigg( 256 H^5 {\dot \xi}^3+256 H^3 {\dot H} {\dot \xi}^3-64 H^4 {\dot \xi}^2 {\ddot \xi}+48 H^4 F {\dot \xi}^2+128 H^3 {\dot F} {\dot \xi}^2+16 H^2 \rew {\dot \varphi}^2 {\dot \xi}^2 \notag \\
&\hspace{40mm}+64 H^2 {\dot H} F {\dot \xi}^2+64 H {\dot H} {\dot F} {\dot \xi}^2-32 H^2 {\dot F} {\dot \xi} {\ddot \xi}+24 H^2 F {\dot F} {\dot \xi}+8H F {\dot \xi} \rew {\dot \varphi}^2+16 H {\dot F}^2 {\dot \xi} \notag \\
&\hspace{40mm}+16 {\dot H} F {\dot F} {\dot \xi}+3 F {\dot F}^2-4 {\dot F}^2 {\ddot \xi}+\rew {\dot \varphi}^2 F^2 \bigg), \label{h11} \\
E_{12}&=\frac{1}{( {\dot F}+2HF+12H^2 {\dot \xi} ) (F_\sigma+4H^2 \xi_\sigma)} \bigg( -16 H^3 {\dot \xi} \xi_\sigma \rew {\dot \varphi}+32 H^2 {\dot H} {\dot \xi} F_\varphi \xi_\sigma-32 H^2 {\dot H} {\dot \xi} F_\sigma \xi_\varphi-4H^2 F\xi_\sigma \rew {\dot \varphi} \notag \\
&\hspace{60mm}-4H {\dot \xi} F_\sigma \rew {\dot \varphi}+8 {\dot H} {\dot F} \xi_\sigma F_\varphi-8 {\dot H} {\dot F} F_\sigma \xi_\varphi-F F_\sigma \rew {\dot \varphi} \bigg), \label{h12} \\
E_{22}&=\frac{8\pi w}{M_P^2}. \label{h22}
\end{align}

\section{Matrix elements of $A_{ab}, E_{ab}$ for the special case}
The matrix elements of $A_{ab}$ are given by
\begin{align}
A_{11}&=\frac{2( F+4H {\dot \xi})}{ {({\dot F}+2HF+12H^2 {\dot \xi})}^2 } \left[ 3{\dot F}^2+24H^2 {\dot F} {\dot \xi}+48 H^4 {\dot \xi}^2+\rew (F+4H {\dot \xi}) {\dot \varphi}^2 \right], \\
A_{12}&=-\frac{F+4H{\dot \xi}}{{\dot F}+2HF+12H^2 {\dot \xi}} \rew {\dot \varphi}, \\
A_{22}&=\frac{8\pi w}{M_P^2}.
\end{align}

As for the matrix elements of $E_{ab}$, they are given by (\ref{h11}), (\ref{h12}) and (\ref{h22}).

We also give expressions for $\det A,~\det E$ and $\Tr A\, \Tr E-\Tr (AE)$,
which are necessary for deriving the dispersion relation.
They are given by
\begin{alignat}{4}
\det A &= \frac{3 \rew (F+4H{\dot \xi}) {({\dot F}+4H^2 {\dot \xi})}^2}{{({\dot F}+2HF+12H^2 {\dot \xi})}^2}, \\
\Tr A \Tr E&=\Tr  (AE)+\frac{2 \rew {({\dot F}+4H^2 {\dot \xi})}^2}{(F_\lambda+4H^2 F_\sigma) {({\dot F}+2HF+12H^2 {\dot \xi})}^2 }
\,(12H^2 F F_\sigma+56 H^3 {\dot \xi} F_\sigma+14 H {\dot \xi}F_\lambda+8 {\dot H} F F_\sigma \notag\\
&-8H^2 {\ddot \xi} F_\sigma-2{\ddot \xi} F_\lambda+3F F_\lambda+32 H {\dot H} {\dot \xi} F_\sigma ), \\
\det E&=-\frac{{({\dot F}+4H^2 {\dot \xi})}^2}{{({\dot F}+2HF+12H^2 {\dot \xi})}^2 {(F_\lambda+4H^2 F_\sigma)}^2 {\dot \varphi}^2} \bigg( 64 {\dot H}^2 {\dot \xi}^2 F_\lambda^2-256 H^3 (H^2+{\dot H}) {\dot \xi} F_\sigma^2 \rew {\dot \varphi}^2 \notag \\
&-128 H^3 {\dot \xi} F_\sigma F_\lambda \rew {\dot \varphi}^2-64 H{\dot H} {\dot \xi} F_\lambda F_\sigma \rew {\dot \varphi}^2-16 H{\dot \xi} F_\lambda^2 \rew {\dot \varphi}^2-128 {\dot H}^2 {\dot F} {\dot \xi} F_\lambda F_\sigma \notag \\
&+64 H^4 {\ddot \xi} F_\sigma^2 \rew {\dot \varphi}^2-48 H^4 F F_\sigma^2 \rew {\dot \varphi}^2+32 H^2 {\ddot \xi} F_\sigma F_\lambda \rew {\dot \varphi}^2-24H^2 F F_\lambda F_\sigma \rew {\dot \varphi}^2 \notag \\
&-64 H^2 {\dot H} F F_\sigma^2 \rew {\dot \varphi}^2+64 {\dot H}^2 {\dot F}^2 F_\sigma^2-3 F F_\lambda^2 \rew {\dot \varphi}^2-16 {\dot H}F F_\lambda F_\sigma \rew {\dot \varphi}^2+4 {\ddot \xi} F_\lambda^2 \rew {\dot \varphi}^2 \bigg).
\end{alignat}

\bibliography{mgmScalbib.bib}

\begin{thebibliography}{28}
\expandafter\ifx\csname natexlab\endcsname\relax\def\natexlab#1{#1}\fi
\expandafter\ifx\csname bibnamefont\endcsname\relax
  \def\bibnamefont#1{#1}\fi
\expandafter\ifx\csname bibfnamefont\endcsname\relax
  \def\bibfnamefont#1{#1}\fi
\expandafter\ifx\csname citenamefont\endcsname\relax
  \def\citenamefont#1{#1}\fi
\expandafter\ifx\csname url\endcsname\relax
  \def\url#1{\texttt{#1}}\fi
\expandafter\ifx\csname urlprefix\endcsname\relax\def\urlprefix{URL }\fi
\providecommand{\bibinfo}[2]{#2}
\providecommand{\eprint}[2][]{\url{#2}}

\bibitem[{\citenamefont{De~Felice and Suyama}(2009)}]{DeFelice:2009ak}
\bibinfo{author}{\bibfnamefont{A.}~\bibnamefont{De~Felice}} \bibnamefont{and}
  \bibinfo{author}{\bibfnamefont{T.}~\bibnamefont{Suyama}},
  \bibinfo{journal}{JCAP} \textbf{\bibinfo{volume}{0906}}, \bibinfo{pages}{034}
  (\bibinfo{year}{2009}), \eprint{0904.2092}.

\bibitem[{\citenamefont{{Carroll} et~al.}(2005)\citenamefont{{Carroll}, {De
  Felice}, {Duvvuri}, {Easson}, {Trodden}, and {Turner}}}]{2005PhRvD..71f3513C}
\bibinfo{author}{\bibfnamefont{S.~M.} \bibnamefont{{Carroll}}},
  \bibinfo{author}{\bibfnamefont{A.}~\bibnamefont{{De Felice}}},
  \bibinfo{author}{\bibfnamefont{V.}~\bibnamefont{{Duvvuri}}},
  \bibinfo{author}{\bibfnamefont{D.~A.} \bibnamefont{{Easson}}},
  \bibinfo{author}{\bibfnamefont{M.}~\bibnamefont{{Trodden}}},
  \bibnamefont{and} \bibinfo{author}{\bibfnamefont{M.~S.}
  \bibnamefont{{Turner}}}, \bibinfo{journal}{\prd}
  \textbf{\bibinfo{volume}{71}}, \bibinfo{pages}{063513}
  (\bibinfo{year}{2005}), \eprint{arXiv:astro-ph/0410031}.

\bibitem[{\citenamefont{Nunez and Solganik}(2005)}]{Nunez:2004ts}
\bibinfo{author}{\bibfnamefont{A.}~\bibnamefont{Nunez}} \bibnamefont{and}
  \bibinfo{author}{\bibfnamefont{S.}~\bibnamefont{Solganik}},
  \bibinfo{journal}{Phys. Lett.} \textbf{\bibinfo{volume}{B608}},
  \bibinfo{pages}{189} (\bibinfo{year}{2005}), \eprint{hep-th/0411102}.

\bibitem[{\citenamefont{Nojiri et~al.}(2007)\citenamefont{Nojiri, Odintsov, and
  Tretyakov}}]{Nojiri:2007te}
\bibinfo{author}{\bibfnamefont{S.}~\bibnamefont{Nojiri}},
  \bibinfo{author}{\bibfnamefont{S.~D.} \bibnamefont{Odintsov}},
  \bibnamefont{and} \bibinfo{author}{\bibfnamefont{P.~V.}
  \bibnamefont{Tretyakov}}, \bibinfo{journal}{Phys. Lett.}
  \textbf{\bibinfo{volume}{B651}}, \bibinfo{pages}{224} (\bibinfo{year}{2007}),
  \eprint{0704.2520}.

\bibitem[{\citenamefont{Navarro and Van~Acoleyen}(2005)}]{Navarro:2005gh}
\bibinfo{author}{\bibfnamefont{I.}~\bibnamefont{Navarro}} \bibnamefont{and}
  \bibinfo{author}{\bibfnamefont{K.}~\bibnamefont{Van~Acoleyen}},
  \bibinfo{journal}{Phys. Lett.} \textbf{\bibinfo{volume}{B622}},
  \bibinfo{pages}{1} (\bibinfo{year}{2005}), \eprint{gr-qc/0506096}.

\bibitem[{\citenamefont{Navarro and Van~Acoleyen}(2006)}]{Navarro:2005da}
\bibinfo{author}{\bibfnamefont{I.}~\bibnamefont{Navarro}} \bibnamefont{and}
  \bibinfo{author}{\bibfnamefont{K.}~\bibnamefont{Van~Acoleyen}},
  \bibinfo{journal}{JCAP} \textbf{\bibinfo{volume}{0603}}, \bibinfo{pages}{008}
  (\bibinfo{year}{2006}), \eprint{gr-qc/0511045}.

\bibitem[{\citenamefont{Mena et~al.}(2006)\citenamefont{Mena, Santiago, and
  Weller}}]{Mena:2005ta}
\bibinfo{author}{\bibfnamefont{O.}~\bibnamefont{Mena}},
  \bibinfo{author}{\bibfnamefont{J.}~\bibnamefont{Santiago}}, \bibnamefont{and}
  \bibinfo{author}{\bibfnamefont{J.}~\bibnamefont{Weller}},
  \bibinfo{journal}{Phys. Rev. Lett.} \textbf{\bibinfo{volume}{96}},
  \bibinfo{pages}{041103} (\bibinfo{year}{2006}), \eprint{astro-ph/0510453}.

\bibitem[{\citenamefont{De~Felice et~al.}(2006)\citenamefont{De~Felice,
  Hindmarsh, and Trodden}}]{DeFelice:2006pg}
\bibinfo{author}{\bibfnamefont{A.}~\bibnamefont{De~Felice}},
  \bibinfo{author}{\bibfnamefont{M.}~\bibnamefont{Hindmarsh}},
  \bibnamefont{and} \bibinfo{author}{\bibfnamefont{M.}~\bibnamefont{Trodden}},
  \bibinfo{journal}{JCAP} \textbf{\bibinfo{volume}{0608}}, \bibinfo{pages}{005}
  (\bibinfo{year}{2006}), \eprint{astro-ph/0604154}.

\bibitem[{\citenamefont{Bonvin et~al.}(2006)\citenamefont{Bonvin, Caprini, and
  Durrer}}]{Bonvin:2006vc}
\bibinfo{author}{\bibfnamefont{C.}~\bibnamefont{Bonvin}},
  \bibinfo{author}{\bibfnamefont{C.}~\bibnamefont{Caprini}}, \bibnamefont{and}
  \bibinfo{author}{\bibfnamefont{R.}~\bibnamefont{Durrer}},
  \bibinfo{journal}{Phys. Rev. Lett.} \textbf{\bibinfo{volume}{97}},
  \bibinfo{pages}{081303} (\bibinfo{year}{2006}), \eprint{astro-ph/0606584}.

\bibitem[{\citenamefont{Hashimoto and Itzhaki}(2001)}]{Hashimoto:2000ys}
\bibinfo{author}{\bibfnamefont{A.}~\bibnamefont{Hashimoto}} \bibnamefont{and}
  \bibinfo{author}{\bibfnamefont{N.}~\bibnamefont{Itzhaki}},
  \bibinfo{journal}{Phys. Rev.} \textbf{\bibinfo{volume}{D63}},
  \bibinfo{pages}{126004} (\bibinfo{year}{2001}), \eprint{hep-th/0012093}.

\bibitem[{\citenamefont{Bruneton}(2007)}]{Bruneton:2006gf}
\bibinfo{author}{\bibfnamefont{J.-P.} \bibnamefont{Bruneton}},
  \bibinfo{journal}{Phys. Rev.} \textbf{\bibinfo{volume}{D75}},
  \bibinfo{pages}{085013} (\bibinfo{year}{2007}), \eprint{gr-qc/0607055}.

\bibitem[{\citenamefont{Ellis et~al.}(2007)\citenamefont{Ellis, Maartens, and
  MacCallum}}]{Ellis:2007ic}
\bibinfo{author}{\bibfnamefont{G.}~\bibnamefont{Ellis}},
  \bibinfo{author}{\bibfnamefont{R.}~\bibnamefont{Maartens}}, \bibnamefont{and}
  \bibinfo{author}{\bibfnamefont{M.~A.~H.} \bibnamefont{MacCallum}},
  \bibinfo{journal}{Gen. Rel. Grav.} \textbf{\bibinfo{volume}{39}},
  \bibinfo{pages}{1651} (\bibinfo{year}{2007}), \eprint{gr-qc/0703121}.

\bibitem[{\citenamefont{Babichev et~al.}(2008)\citenamefont{Babichev, Mukhanov,
  and Vikman}}]{Babichev:2007dw}
\bibinfo{author}{\bibfnamefont{E.}~\bibnamefont{Babichev}},
  \bibinfo{author}{\bibfnamefont{V.}~\bibnamefont{Mukhanov}}, \bibnamefont{and}
  \bibinfo{author}{\bibfnamefont{A.}~\bibnamefont{Vikman}},
  \bibinfo{journal}{JHEP} \textbf{\bibinfo{volume}{02}}, \bibinfo{pages}{101}
  (\bibinfo{year}{2008}), \eprint{0708.0561}.

\bibitem[{\citenamefont{Starobinsky}(1980)}]{Starobinsky:1980te}
\bibinfo{author}{\bibfnamefont{A.~A.} \bibnamefont{Starobinsky}},
  \bibinfo{journal}{Phys. Lett.} \textbf{\bibinfo{volume}{B91}},
  \bibinfo{pages}{99} (\bibinfo{year}{1980}).

\bibitem[{\citenamefont{Capozziello}(2002)}]{Capozziello:2002rd}
\bibinfo{author}{\bibfnamefont{S.}~\bibnamefont{Capozziello}},
  \bibinfo{journal}{Int. J. Mod. Phys.} \textbf{\bibinfo{volume}{D11}},
  \bibinfo{pages}{483} (\bibinfo{year}{2002}), \eprint{gr-qc/0201033}.

\bibitem[{\citenamefont{Carroll et~al.}(2004)\citenamefont{Carroll, Duvvuri,
  Trodden, and Turner}}]{Carroll:2003wy}
\bibinfo{author}{\bibfnamefont{S.~M.} \bibnamefont{Carroll}},
  \bibinfo{author}{\bibfnamefont{V.}~\bibnamefont{Duvvuri}},
  \bibinfo{author}{\bibfnamefont{M.}~\bibnamefont{Trodden}}, \bibnamefont{and}
  \bibinfo{author}{\bibfnamefont{M.~S.} \bibnamefont{Turner}},
  \bibinfo{journal}{Phys. Rev.} \textbf{\bibinfo{volume}{D70}},
  \bibinfo{pages}{043528} (\bibinfo{year}{2004}), \eprint{astro-ph/0306438}.

\bibitem[{\citenamefont{Brans and Dicke}(1961)}]{Brans:1961sx}
\bibinfo{author}{\bibfnamefont{C.}~\bibnamefont{Brans}} \bibnamefont{and}
  \bibinfo{author}{\bibfnamefont{R.~H.} \bibnamefont{Dicke}},
  \bibinfo{journal}{Phys. Rev.} \textbf{\bibinfo{volume}{124}},
  \bibinfo{pages}{925} (\bibinfo{year}{1961}).

\bibitem[{\citenamefont{Dicke}(1962)}]{Dicke:1961gz}
\bibinfo{author}{\bibfnamefont{R.~H.} \bibnamefont{Dicke}},
  \bibinfo{journal}{Phys. Rev.} \textbf{\bibinfo{volume}{125}},
  \bibinfo{pages}{2163} (\bibinfo{year}{1962}).

\bibitem[{\citenamefont{Hwang and Noh}(2005)}]{Hwang:2005hb}
\bibinfo{author}{\bibfnamefont{J.-c.} \bibnamefont{Hwang}} \bibnamefont{and}
  \bibinfo{author}{\bibfnamefont{H.}~\bibnamefont{Noh}},
  \bibinfo{journal}{Phys. Rev.} \textbf{\bibinfo{volume}{D71}},
  \bibinfo{pages}{063536} (\bibinfo{year}{2005}), \eprint{gr-qc/0412126}.

\bibitem[{\citenamefont{De~Felice and
  Tsujikawa}(2009{\natexlab{a}})}]{DeFelice:2008wz}
\bibinfo{author}{\bibfnamefont{A.}~\bibnamefont{De~Felice}} \bibnamefont{and}
  \bibinfo{author}{\bibfnamefont{S.}~\bibnamefont{Tsujikawa}},
  \bibinfo{journal}{Phys. Lett.} \textbf{\bibinfo{volume}{B675}},
  \bibinfo{pages}{1} (\bibinfo{year}{2009}{\natexlab{a}}), \eprint{0810.5712}.

\bibitem[{\citenamefont{De~Felice and
  Tsujikawa}(2009{\natexlab{b}})}]{DeFelice:2009aj}
\bibinfo{author}{\bibfnamefont{A.}~\bibnamefont{De~Felice}} \bibnamefont{and}
  \bibinfo{author}{\bibfnamefont{S.}~\bibnamefont{Tsujikawa}}
  (\bibinfo{year}{2009}{\natexlab{b}}), \eprint{0907.1830}.

\bibitem[{\citenamefont{Traschen and Brandenberger}(1990)}]{Traschen:1990sw}
\bibinfo{author}{\bibfnamefont{J.~H.} \bibnamefont{Traschen}} \bibnamefont{and}
  \bibinfo{author}{\bibfnamefont{R.~H.} \bibnamefont{Brandenberger}},
  \bibinfo{journal}{Phys. Rev.} \textbf{\bibinfo{volume}{D42}},
  \bibinfo{pages}{2491} (\bibinfo{year}{1990}).

\bibitem[{\citenamefont{Kofman et~al.}(1994)\citenamefont{Kofman, Linde, and
  Starobinsky}}]{Kofman:1994rk}
\bibinfo{author}{\bibfnamefont{L.}~\bibnamefont{Kofman}},
  \bibinfo{author}{\bibfnamefont{A.~D.} \bibnamefont{Linde}}, \bibnamefont{and}
  \bibinfo{author}{\bibfnamefont{A.~A.} \bibnamefont{Starobinsky}},
  \bibinfo{journal}{Phys. Rev. Lett.} \textbf{\bibinfo{volume}{73}},
  \bibinfo{pages}{3195} (\bibinfo{year}{1994}), \eprint{hep-th/9405187}.

\bibitem[{\citenamefont{Groot~Nibbelink and van
  Tent}(2002)}]{GrootNibbelink:2001qt}
\bibinfo{author}{\bibfnamefont{S.}~\bibnamefont{Groot~Nibbelink}}
  \bibnamefont{and} \bibinfo{author}{\bibfnamefont{B.~J.~W.} \bibnamefont{van
  Tent}}, \bibinfo{journal}{Class. Quant. Grav.} \textbf{\bibinfo{volume}{19}},
  \bibinfo{pages}{613} (\bibinfo{year}{2002}), \eprint{hep-ph/0107272}.

\bibitem[{\citenamefont{Rebane}(1969)}]{Rebane:1969}
\bibinfo{author}{\bibfnamefont{T.~K.} \bibnamefont{Rebane}},
  \bibinfo{journal}{Theoretical and Experimental Chemistry}
  \textbf{\bibinfo{volume}{5}}, \bibinfo{pages}{1} (\bibinfo{year}{1969}).

\bibitem[{\citenamefont{Ternov et~al.}(1971)\citenamefont{Ternov, Bagrov, and
  Zadorozhnyi}}]{Ternov:1970}
\bibinfo{author}{\bibfnamefont{I.~M.} \bibnamefont{Ternov}},
  \bibinfo{author}{\bibfnamefont{V.~G.} \bibnamefont{Bagrov}},
  \bibnamefont{and} \bibinfo{author}{\bibfnamefont{V.~N.}
  \bibnamefont{Zadorozhnyi}}, \bibinfo{journal}{Soviet Physics Journal}
  \textbf{\bibinfo{volume}{14}}, \bibinfo{pages}{492} (\bibinfo{year}{1971}).

\bibitem[{\citenamefont{Abdalla}(1988{\natexlab{a}})}]{Abdalla:1988a}
\bibinfo{author}{\bibfnamefont{M.~S.} \bibnamefont{Abdalla}},
  \bibinfo{journal}{Phys. Rev. A} \textbf{\bibinfo{volume}{37}},
  \bibinfo{pages}{4026} (\bibinfo{year}{1988}{\natexlab{a}}).

\bibitem[{\citenamefont{Abdalla}(1988{\natexlab{b}})}]{Abdalla:1988b}
\bibinfo{author}{\bibfnamefont{M.~S.} \bibnamefont{Abdalla}},
  \bibinfo{journal}{Nuovo Cim. B} \textbf{\bibinfo{volume}{101}},
  \bibinfo{pages}{267} (\bibinfo{year}{1988}{\natexlab{b}}).

\end{thebibliography}

\end{document}